\documentclass[final,3p,times,twocolumn]{elsarticle}
 \biboptions{comma,sort&compress}
\usepackage{here}
 \usepackage{graphicx}
\usepackage{epsfig}
\usepackage{amsmath}
\usepackage{latexsym}
\usepackage{stmaryrd}

\newcommand{\bear}{\begin{eqnarray}}
\newcommand{\eear}{\end{eqnarray}}
  
\def\nin{\noindent}
\def\beq{\begin{equation}}
\def\eeq{\end{equation}}
\def\bea{\begin{eqnarray}}
\def\eea{\end{eqnarray}}

\journal{Nuc. Phys. (Proc. Suppl.)}

\begin{document}

\begin{frontmatter}



\title{Extracting gluon condensate from the average plaquette}

 \author{Taekoon Lee}
  \address{Department of Physics, Kunsan National University, 
Kunsan 573-701, Korea}
\ead{tlee@kunsan.ac.kr}



\begin{abstract}
\noindent{}
The perturbative contribution in the average plaquette is
subtracted using Borel summation and the remnant of the plaquette 
is shown to scale
  as a dim-4 condensate. 
A critical review is presented  of the 
renormalon subtraction scheme that claimed a dim-2 condensate.
The extracted gluon condensate is compared with the latest result
employing high order (35-loop) calculation in the
 stochastic perturbation theory. 
\end{abstract}

\begin{keyword}
renormalon \sep plaquette \sep gluon condensate
 \sep Borel summation \sep bilocal expansion


\end{keyword}

\end{frontmatter}


\section{Introduction}
\nin
Extracting
 the gluon condensate from the average plaquette is an old problem of
 lattice gauge theory. The operator product expansion (OPE) for the
 plaquette in pure ${\rm SU}(3)$
  Yang-Mills theory is given by
\bear
P(\beta)\!\!&\equiv&\!\!\langle 1- 
\frac{1}{3}\text{Tr} \,\text{U}_{\boxempty}
\rangle \nonumber \\
\!\!&=&\!\!\sum_{n=1}\frac{c_n}{\beta^n} + \frac{\pi^2}{36}
Z(\beta)\langle\frac{\alpha_s}{\pi}GG \rangle a^4 +O(a^6) \,,
\label{ope}
\eear
where $\beta$ is the lattice coupling and $a$ the lattice 
spacing.

To extract the gluon condensate the perturbative contribution, which
 dominates the plaquette, must be subtracted accurately, to 
 better than one part in $10^4$. 
The perturbative series is expected to be an asymptotic series and at high
orders the coefficients to be dominated 
by renormalon-caused large order behavior. 
 However, the perturbative coefficients computed 
using stochastic perturbation theory \cite{direnzo2} to 10-loop order does
not display a renormalon behavior but a power law that grows much faster 
than the expected renormalon-caused large order behavior.

To subtract the perturbative contribution Burgio et al. \cite{direnzo} introduced a 
continuum scheme in which the renormalon contribution
 to the perturbative part of the plaquette is calculated by matching
  the computed high order coefficients 
   with the renormalon-caused large order
    behavior mapped to the lattice scheme. This scheme of subtracting
    perturbative contribution resulted in a
 power correction that scales, surprisingly, as a dim-2 
condensate. This is in contradiction with 
the OPE~(\ref{ope}) that demands 
the leading power correction be of a dim-4 condensate.

In this presentation based on \cite{lee.condensate} we give  a critical
 review of the above  renormalon subtraction scheme  and show that
  the continuum scheme in which the renormalon was subtracted is
 far from a renormalon-dominated scheme and it fails 
 as well the self-consistency test.

We then introduce a renormalon subtraction scheme based on the bilocal
expansion of the Borel transform in continuum scheme, and show that
 the plaquette data less the Borel-summed 
perturbative contribution scales as a dim-4 condensate.
The extracted gluon condensate is then compared with the result from the
recent  high order (35-loop) calculation of 
the plaquette using stochastic perturbation theory.

\section{Critical review of existing renormalon subtraction scheme}
\nin
The renormalon subtraction scheme of Burgio et at.~\cite{direnzo}
  writes essentially  the
 average plaquette as
\bear
P(\beta)=P^{\rm ren}(\beta_c)+ \delta P(\beta_c) +P_{\rm NP}(\beta)\,,
\label{decomposition}
\eear
where 
\bear
P^{\rm ren}(\beta_c)= \int_0^{b_{\rm max}} e^{-\beta_c b}
\frac{\cal N}{(1-b/z_0)^{1+\nu}} db \eear
with $\beta_c$ denoting the coupling in the continuum scheme defined by    
\bear
\beta_c=\beta-r_1-\frac{r_2}{\beta}
\label{beta_rel}
\eear    
and
\bear
z_0=\frac{16\pi^2}{33}\,,\quad \nu=\frac{204}{121}\,.
\label{consts}
\eear
Here the plaquette is divided into 
the renormalon contribution  $P^{\rm 
ren}$ and the rest of the perturbative contribution  $\delta P$,   and 
nonperturbative correction $P_{\rm NP}$. 
The asymptotically divergent behavior
of the perturbative series is contained in $P^{\rm 
ren}$, and  $\delta P$ denotes the rest that can be
 expressed as a convergent series.

Now define $ P_{\rm NP}^{(N)}$ with
\bear
P_{\rm NP}^{(N)}(\beta)\equiv P(\beta)-
P^{\rm ren}(\beta_c)-\sum_{n=1}^{N}
(c_n-C_n^{\rm ren})\beta^{-n}\,,
\label{powercorrection}
\eear
where $C_n^{\rm ren}$ denotes the perturbative coefficients in the
lattice scheme of
$P^{\rm ren}$. Note that, by definition,
$P_{\rm NP}^{(N)}$ is free of perturbative coefficients to order $N$.
The constants $r_1,r_2$ that define the continuum scheme and the
normalization constant $\cal N$  are to be determined 
so that    $C_n^{\rm
ren}$ converges to $c_n$ at large orders. In the continuum scheme given by
\bear
r_1=3.1\,, \quad r_2=2.0\,,
\label{scheme}
\eear
and with an appropriate value for $\cal N$ it was observed that 
 $C_n^{\rm ren}$ converge to $c_n$ at the orders computed in 
stochastic
perturbation theory. Because the last term in 
(\ref{powercorrection})  is a 
convergent series  $ P_{\rm NP}^{(N)}$
is well-defined at $N\to\infty$, and this is precisely the  
quantity that was assumed
to represent the power correction, and it was $P_{\rm NP}^{(8)}$ that 
was shown to scale as a dim-2 condensate.

In this procedure  the large order 
behaviors in the
 lattice scheme and the continuum scheme are matched  using
 only the high order
 coefficients. However, this matching of large order behaviors
 would only work when the perturbative coefficients in both schemes
  are already in
 asymptotic regime and 
 follow the pattern of 
renormalon-caused large order behavior.
But, because the computed coefficients in the lattice scheme are 
far from being in the asymptotic regime  
and do not follow the renormalon pattern
the matching cannot be performed. Therefore, the conclusion 
of a dimension-2 condensate based on this matching 
should be reexamined.

One way to check the consistency of the above procedure is to map
the computed coefficients in the lattice scheme to the continuum scheme and
see if the mapped coefficients follow the pattern of a renormalon behavior.
As can be seen in Table ~\ref{Table1}, however, the coefficients in the
continuum scheme are sign-alternating, instead of a factorially growing
pattern expected from a  renormalon behavior.
 It shows 
that the  mapping of the
perturbative coefficients between the 
lattice scheme and the continuum scheme of 
(\ref{scheme}) at the orders in consideration
 are still very sensitive
on the low order coefficients, 
which were ignored in the matching procedure of \cite{direnzo}.
It is thus obvious that the continuum scheme of  (\ref{scheme})
cannot be the right scheme where renormalon can be subtracted reliably.
 
 {\scriptsize
\begin{table}[thb]
\setlength{\tabcolsep}{1.2pc}
 \caption{\scriptsize The perturbative coefficients 
of the average plaquette in the
continuum scheme.   }
    {\small
\begin{tabular}{cccc}
	\hline
$c_1^{\text{cont}}$&$c_2^{\text{cont}}$&$c_3^{\text{cont}}$&
$c_4^{\text{cont}}$\\
2.0&-4.9792&10.613&-10.200\\ \hline

$c_5^{\text{cont}}$&$c_6^{\text{cont}}$&
$c_7^{\text{cont}}$&$c_8^{\text{cont}}$  \\ 
-44.218&316.34&-1096.&1947.\\ \hline
\end{tabular}
}
\label{Table1}
\end{table}
}

Checking the internal consistency of the 
subtraction scheme
also shows an underlying problem. 
The nonperturbative term in (\ref{decomposition}) can be written 
using  (\ref{powercorrection}) as
\bear
\!\!\!\!P_{\rm NP}(\beta)\!=\!P_{\rm NP}^{(N)}(\beta) \!-\!\left[\delta 
P(\beta_c)\!-\!\sum_{n=1}^N(c_n\!-\!C_n^{\rm
ren}) \beta^{-n}\right]\,.
\eear 
For $  P_{\rm NP}^{(N)}$ to approximate the power correction 
\bear
\left|\delta P(\beta_c)-\sum_{n=1}^N(c_n-C_n^{\rm ren}) 
\beta^{-n}\right| \ll
  P_{\rm NP}^{(N)} (\beta)
\label{criterion}
\eear
must be satisfied. Because $\delta P(\beta_c)$ is by definition 
a convergent
quantity it can be written in a series expansion
 \bear
\delta P(\beta_c)\equiv \sum_{n=1}^\infty D_n \beta_c^{-n}\,,
\eear{}
where $D_n$ can be computed to the order $c_n$ are 
known, and
 (\ref{criterion}) can be written approximately   as 
\bear
\frac{|\sum_{n=1}^N D_n \beta_c^{-n}-\sum_{n=1}^N(c_n-C_n^{\rm ren})
\beta^{-n}|}{ P_{\rm NP}^{(N)}(\beta)}\ll1 \,.
\label{criterion1}
\eear
In the continuum scheme of  (\ref{scheme}), and  at $N=8$ and 
$\beta=6.0, 6.2$ and  6.4, for example,  the ratios are $69, 59$ and 
42, respectively, which is a 
 severe violation of the consistency condition. This again
confirms  that the continuum 
scheme of (\ref{scheme}) cannot be a proper scheme for 
renormalon  subtraction.


\nin{}
\section{Renormalon subtraction by Borel summation}
\nin{}
Clearly, the perturbative contribution 
in the plaquette cannot be subtracted by matching the
renormalon-based coefficients in a continuum
scheme to the lattice scheme. Instead,  one must map the
 perturbative coefficients in the lattice scheme to a continuum
 scheme and search for a scheme in which the mapped coefficients
 display a renormalon pattern.
Once such a scheme is found one can then use
Borel summation in that scheme to subtract the
perturbative contribution to extract the nonperturbative
 power correction.

In this paper we shall assume that such a scheme exists and 
perform  Borel summation  using the  bilocal
expansion of Borel transform \cite{surviving}. To Borel-sum the 
divergent perturbative series accurately, one must have a 
precise description of the Borel transform in the domain
that contains the origin as well as the first renormalon singularity in 
the Borel plane. The bilocal expansion, utilizing the known perturbative
coefficients and the properties of 
the first renormalon singularity, reconstructs the Borel transform in the
above domain of interest by interpolating the expansions about the origin
and  about the renormalon singularity. The Borel transform 
in the form of bilocal expansion implements the correct
 nature of the first renormalon singularity as well
 as the perturbative coefficients that match
the known coefficients.
The sum of the
Borel-summed perturbative contribution and a dim-4 power correction
 is then  fitted
to the plaquette data. A good fit would then suggest 
the power correction be of dim-4 type.

The  Borel summation using bilocal expansion in the continuum scheme
 using the first $N$-loop perturbations 
of the plaquette 
is given in the form:
\bear
\!\!\!\!\!\!\!\!\!\!\!{P}_{\rm BR}^{\rm (N)}(\beta) \!={}
\!{\rm Re}\!\!\!\int_{0}^{\infty}\!\!\!{}
 e^{-\beta_c b}
\left[\sum_{n=0}^{N-1} \frac{h_n}{n!} 
b^n\!+\!\frac{\cal N}{(1-b/z_0)^{1+\nu}} \right] db\,,
\label{bilocal}
\eear
where the integration  is performed along
 the real axis on the upper-half plane.
 The essential idea of the
bilocal expansion is to interpolate
the two perturbative expansions 
about the origin and about the renormalon singularity to
rebuild the Borel transform. Incorporating the renormalon singularity
explicitly extends  the applicable domain of the Borel transform from
near the origin to areas
beyond the renormalon singularity. 
This scheme was used in summing the perturbative series for
 the static inter-quark potential 
as well as the  heavy quark pole mass \cite{surviving,heavyquark};
 The Borel-summed static potential agrees very well with lattice results and
 the convergence in the pole mass case is extremely fast.

${\cal N}$ in (\ref{bilocal}) denotes the normalization constant 
of the large order behavior
and the coefficients $h_n$ are determined so that the Borel transform
 reproduce the 
perturbative coefficients in the continuum scheme when
expanded about the origin.
Thus $h_n$ depends on the continuum perturbative coefficients 
as well as ${\cal N}$. By definition, ${P}_{\rm BR}^{(N)}(\beta)$, when
expanded in $1/\beta$, reproduces the 
perturbative coefficients of the average 
plaquette to the $N$-loop order that were 
employed in building the Borel transform. 
The power correction can  then be defined by:
\bear
{ P}_{\rm NP}^{\rm (N)}(\beta)\equiv {P}(\beta) -
{P}_{\rm BR}^{\rm (N)}(\beta)\,,
\label{pNP}
\eear
which, by definition, is free of perturbative coefficients
 to order $N$.

Using the perturbation to 10-loop order of the plaquette 
we compute ${P}_{\rm BR}^{(10)}(\beta)$ in the continuum scheme 
defined  by (\ref{beta_rel}). 
The normalization $\cal{N}$ is treated as a
 fitting parameter, and  in our scheme 
the parameters to be fitted are  $r_1, r_2$, and $\cal{N}$.

\begin{figure}[hbt] 
\centerline{\includegraphics[width=6.cm]{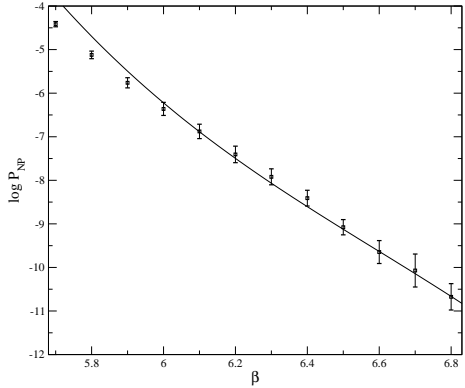}}
\caption{\scriptsize $\log { P}_{\rm NP}$ vs. $\beta$. 
The solid line is for $4 \log(a/r_0) +0.5$. 
The plot shows the power 
correction should be of dim-4 type.
}
\label{fig1} 
\end{figure}

 Using the plaquette data for 
$6.0\leq \beta \leq 6.8$ from~\cite{plaquette} and
the relation between the lattice spacing $a$ 
and $\beta$ from static quark force 
simulation \cite{sommer}
\bear
\log(a/r_0)&\!\!\!=\!\!\!&-1.6804 - 
1.7331(\beta - 6) + \nonumber \\
\!\!\!&&\!\!\! 0.7849(\beta - 6)^2 - 0.4428(\beta - 6)^3
 \eear
 the fit gives ${\cal N}=165$ and
 \bear
r_1=1.611, \quad r_2=0.246\,,
\label{fitted}
\eear
which values are substantially 
different from those in (\ref{scheme}).
 The result of the fit is shown in 
 Fig. \ref{fig1}, which shows that the power correction is 
 consistent with 
 a dim-4 condensate. The agreement improves as $\beta$
  increases, albeit with larger uncertainties;
   The  deviation at low $\beta$ ($\beta <6$)
   may be attributed to a dim-6
 condensate, which may be seen, though not presented here, 
  by that adding a 
 dim-6 power correction in the fit  
 improves the agreement in the whole
 range of the plot.   The error bars are from the 
 uncertainty in the simulated
 perturbative coefficients of the plaquette. 
 The uncertainty in the normalization constant
 does not appear to be large: for example, 
 a variation of 20\% in ${\cal N}$ causes 
 less than a quarter of those by the perturbative coefficients.

From the fit we obtain a dim-4 power correction of
 $P_{\rm NP}\approx 1.6\,\, (a/r_0)^4$. 
 Because of the asymptotic nature of the perturbative series 
 the power correction of the plaquette 
 is dependent on the subtraction scheme
 of the perturbative contribution, and 
 thus our result may not be 
 directly compared to those from other 
 subtraction schemes. Nevertheless, it is 
 still interesting to observe that 
the result is roughly consistent 
with $0.4\,\, (a/r_0)^4$ of  \cite{rakow} 
and $0.7 \,\,(a/r_0)^4$ of  \cite{meurice}. 
Our result turns out to be
a little  larger
than those estimates; This may be partly accounted for
by the fact that the existing 
results were from the fit in the low $\beta$ range
of $\beta \lesssim 6$, in which 
range the data are below our fitted curve.
\nin

\section{Comparison with  latest
 result from 35-loop calculation}
 \nin{}
 Recently Bali et al. \cite{bali1,bali2}
computed the plaquette to 35-loop order
 using stochastic perturbation theory and observed a renormalon
 behavior of the coefficients at high orders, determining the normalization
 of the large order behavior. With the perturbative expansion
 in the asymptotic regime the power correction
 may as well be extracted by
 subtracting from the average plaquette the perturbative 
 series  truncated at the order the loop correction
  becomes minimal. The power correction thus obtained was shown 
  to be of dim-4 in accordance with the OPE (\ref{ope}), and  
   the gluon condensate was determined to be
   \bear{}
   \langle\frac{\alpha_s}{\pi}GG \rangle
\approx0.077\,\,{\rm GeV}^{4}\,{}
\label{b.con}
\eear{}
with an intrinsic uncertainty of $0.087\,\,{\rm GeV}^{4}$.
The normalization of the large order behavior in the lattice scheme
 was obtained as
\bear{}
{\cal N}_{\rm lat}=4.2(\pm 1.7) \times 10^{5}\,.
\label{b.norm}
\eear{}

  It is  interesting  to compare these numbers with our results.
  From the fit in the previous section we obtain the gluon condensate:
\bear{}
\langle\frac{\alpha_s}{\pi}GG \rangle=\frac{36}{\pi^{2}}
e^{0.5}r_{0}^{-4}
\approx0.14 \,\,{\rm GeV}^{4}\,,{}
\label{con}
\eear{}
and the normalization ${\cal N}=165 \pm 50$ (30\% uncertainty)
 in the continuum scheme
 corresponds to
\bear{}
{\cal N}_{\rm lat}= 
\frac{2\pi}{3}e^{z_{0}r_{1}}{\cal N}=7.6(\pm 2.3)\times 10^{5}\,,
\eear{}
which is, remarkably, consistent with (\ref{b.norm}).
Comparing the gluon condensates (\ref{b.con}) and (\ref{con}) we see that
they are in agreement within the intrinsic uncertainty. The gluon condensate
is dependent on the prescription for the perturbative contribution
 in the plaquette,  and the
difference between the two condensate 
values may be because they came from different prescriptions:
Borel summation and truncated 
power series, respectively.

\section{Conclusions}
\nin
The renormalon subtraction 
procedure of \cite{direnzo} that led to a dim-2 condensate in the plaquette
was reexamined. It is found that the continuum scheme
 employed in the procedure is far from a renormalon-dominated scheme and
 the procedure also fails a consistency check. 
 As a consequence the power correction extracted is severely contaminated
 by perturbative contribution, to discredit the claimed dim-2 condensate.

We then introduced a renormalon subtraction scheme that avoids the problems,
in which the perturbative contribution is obtained by Borel-summing the
perturbative series
in a continuum scheme, employing the rebuilt Borel transform in the framework of 
the bilocal expansion. The power correction obtained in this procedure
is of dim-4, in accordance with the OPE of the plaquette. 
The normalization of the large order behavior of the plaquette
  as well as 
  the gluon condensate extracted are shown to be  in agreement
 with the latest results from 35-loop order calculations.

\section*{Acknowledgements}
\nin
I am thankful to S. Han and G. Bali for useful conversations.
This research was supported by Basic Science
Research Program through the National Research Foundation of Korea
(NRF), funded by the Ministry of Education, Science, and Technology
(2012R1A1A2044543).














\end{document}